\documentclass[12pt]{article}
\begin{document}
\title{Probing the isospin dependent mean field and nucleon nucleon cross
section in the medium by the nucleon emissions }
\author{\small Jian-Ye Liu,Yong-Zhong Xing,Wen-Jun Guo}
\date{}
\maketitle
\begin{center}
$^{1}${\small Institute for the theory of modern physics, Tianshui Normal University, Gansu,
Tianshui 741000, P. R. China}\\
$^{2}${\small Center of Theoretical Nuclear Physics, National Laboratory of
Heavy Ion Accelerator}\\
{\small Lanzhou 730000, P.R. China}\\
$^{3}${\small Institute of Modern Physics, Chinese Academy of Sciences, P.O.Box 31}\\
{\small Lanzhou 730000, P.R. China}\\
$^{4}${\small CCAST(Word Lab.),P.O.Box 8730,Beijing 100080}\\
{\small Lanzhou 730000, P.R. China}\\
\date{}
\maketitle
\begin{minipage}{140mm}
\baselineskip 0.2in \hskip 0.2in We study the isospin effects of
the mean field and two-body collision on the nucleon emissions at
the intermediate energy heavy ion collisions by using an isospin
dependent transport theory. The calculated results show that the
nucleon emission number $N_{n}$ depends sensitively the isospin
effect of nucleon nucleon cross section and weakly on the isospin
dependent mean field for neutron-poor system in higher beam energy
region . In particular ,the correlation between the medium
correction of two-body collision and the momentum dependent
interaction enhances the dependence of nucleon emission number
$N_{n}$ on the isospin effect of nucleon nucleon cross section.
 On the contrary, the ratio of the neutron proton ratio of the
gas phase to the neutron proton ratio of the liquid phase ,i.e.,
the degree of isospin fractionation
$<(N/Z)_{gas}>_{b}/<(N/Z)_{liq}>_{b}$ depends sensitively on the
isospin dependent mean field and weakly on the isospin effect of
two-body collision  for neutron-rich system in the lower beam
energy region. In this case,  $N_{n}$ and
$<(N/Z)_{gas}>_{b}/<(N/Z)_{liq}>_{b}$ are the probes for
extracting the information about the isospin dependent nucleon
nucleon cross section in the medium and the isospin dependent
mean field,respectively.\\\
{\bf PACS number(s)}: 25$\cdot$70$\cdot$Pq\\
\end{minipage}
\end{center}

Correspond auther: E-mail address: Liujy@ns.lzb.ac.cn\\
Phone number:86-0931-4969318(O),8272215(H). Fax number:86-0931-4969201

\newpage
\baselineskip 0.3in
One can study the isospin effects of heavy ion collision to get
the information about isospin dependent in-medium nucleon-nucleon
cross section and symmetry potential. In order to obtain this
information several interesting isospin effects in heavy ion
collisions have been explored both experimentally and
theoretically over the last few years${[1-16,26-28]}$.  For
example, recently, Bao-An Li et al investigated the isospin effect
of the mean field and showed that the neutron-proton ratio of
preequilibrium nucleon emission,the neutron-proton differential
flow and proton ellipse flow are the probes for extracting the
isospin dependent mean field at lower beam energy region
${[1,2,14,15,16]}$. R.Pak et al have found that the isospin
dependence of collective flow and balance energy is mainly
originated from the isospin-dependent in-medium nucleon-nucleon
cross section ${[17,18]}$.  However two essential ingredients in
heavy ion collision dynamics,the symmetry potential and the
isospin dependent in-medium nucleon-nucleon cross section have not
been well determined so far. Recently Bao-An Li ,M. Colonna , M.Di
Toro and V.Baran et al studied the isospin fractionation in the
intermediate energy heavy ion collisions in recent years
${[15,16,19,30]}$. The isospin fractionation is an unequal
partitioning for the neutron to proton ratio N/Z of unstable
asymmetry nuclear matter between low and high density regions. But
the neutron-proton ratios in gas phase and liquid phase calculated
by Bao An Li were from low densities less (free) and higher
(bound)than $\rho_{c}$=1/8$\rho_{0}$ respectively because IBUU
model can not calculate the formation of fragments${[14]}$. Even
though Bao An Li 's work proposed an useful physical  point of review
but it can not compare with the experimental data directly. An
indications of this phenomenon have been found recently in
intermediate energy heavy-ion experiments by H. Xu et al ${[9]}$.
However they did not compare the calculated results with
experimental data directly and there were some assumptions in
their calculations for comparing with the experimental data. In
this case, how to quantify the gas phase and the liquid phase in
isospin fractionation process for comparing with the experimental
data directly is important. Based on the isospin dependent quantum
molecular dynamics model (IQMD)${[2,20,21,22,26]}$ we investigated
the isospin effects of the nucleon emission number $N_{n}$ for
neutron-poor systems in higher beam energy region and quantify the
degree of isospin fractionation,$(N/Z)_{gas}$ / $(N/Z)_{liq}$ by 
using the neutron proton ratios of the gas phase and liquid phase 
for the neutron-rich systems in lower beam energy region. Where
$(N/Z)_{gas}$ and $(N/Z)_{liq}$ are the neutron-proton ratio of
nucleon emission and that of the fragment emission respectively.
The calculated results show that (1)the nucleon emission number
$N_{n}$ depends sensitively on the isospin effect of nucleon
nucleon cross section and weakly on the symmetry potential in the
higher beam energies ,on the contrary, the degree of isospin
fractionation $(N/Z)_{gas}$ / $(N/Z)_{liq}$ is sensitive to the
the symmetry potential and insensitive to the isospin effect of in-medium
nucleon nucleon cross section. It is worth mentioning that our
calculated results can be compared with the experimental data
directly to get the information on the symmetry potential or the
nucleon nucleon
cross section in the medium. \\\
\hskip 0.3in A quantum molecular dynamics (QMD) model [20,21,22]
contains two dynamical ingredients,the density dependent mean
field and the in-medium nucleon nucleon cross section. In order to
describe the isospin dependence appropriately, the QMD model must
be modified properly. The density dependent mean field should
contain correct isospin terms, including the symmetry potential
and the Coulomb potential. The in-medium nucleon nucleon cross
section should be different for neutron neutron(or proton proton )
and neutron proton collisions, in which Pauli blocking should be
included by distinguishing neutrons and protons. In addition, the
initial condition of the
ground state of two colliding nuclei must also contain isospin effects.\\\
 Considering the above ingredients, we have made important modifications
in QMD to obtain an isospin dependent quantum molecular dynamics
(IQMD) $[2,26]$. The initial density distributions of the
colliding nuclei in IQMD are obtained from the calculations of the
Skyrme-Hatree-Fock with parameter set SKM$^{*}$ $[23]$, and the
initial code of IQMD is used to determine the ground state
properties of the colliding nuclei, such as the binding energies
and RMS radii, which agree with the experimental data for
obtaining the parameters of interaction potential as an input data
of the dynamical calculation by using IQMD. The interaction
potential is given by
\begin{equation}
U(\rho)=U^{Sky}+U^{Coul}+U^{sym}+U^{Yuk}+U^{MDI}+U^{Pauli}
\end{equation}
,where $U^{Coul}$ ,$U^{Sky}$, $U^{Yuk}$,$U^{MDI}$ and $U^{Pauli}$ are
the Coulomb potential, the density dependent Skyrme potential, the
Yukawa potential, the momentum dependent interaction  and the
Pauli potential $[31,32]$ respectively ( see in more detail in Refs $[2,20,21,22]$).\\\
In the present paper,three different forms of $U^{sym}$ have been
used ${[1,15]}$,
\begin{equation}
U_1^{sym}=cF_1(u)\delta \tau _z
\end{equation}
\begin{equation}
U_2^{sym}=cF_2(u)\delta \tau _z+\frac 12cF_2(u)\delta ^2
\end{equation}
\begin{equation}
U_3^{sym}=cF_3(u)[\delta \tau _z-\frac {1}{4}\delta ^2]
\end{equation}
\begin{equation}
U_0^{sym}=0.0
\end{equation}

 with
    \[\tau_{z}=\left\{ \begin{array}{ll}
              1 & \mbox{for neutron}\\
             -1 & \mbox{for proton}
             \end{array}
            \right. \]
 Here c is the strength of symmetry potential, taking the value of 32MeV, F$_1
$(u)=u , F$_2$(u)=u$^2$ and $F_3(u)=u^{1/2}$,u$\equiv \frac \rho
{\rho
_0}, \delta $ is the relative neutron excess $\delta =\frac{\rho _n-\rho _p}{%
\rho _n+\rho _p}=\frac{\rho _n-\rho _p}\rho $.$\rho $,$\rho _{_0}$,$%
\rho _n$ and $\rho _p$ are the total , normal , neutron and proton
densities, respectively. Where $U_0^{sym}$=0.0 indicates no
symmetry potential. It is worth mentioning that the recent studies
of collective flow in heavy ion collisions at intermediate energy
have indicated a reduction of in-medium nucleon nucleon cross
sections. An empirical density dependent nucleon nucleon cross
section in medium $[24]$ has been suggested as follows
\begin{equation}
\sigma_{NN}=(1+\alpha\frac{\rho}{\rho_{0}})\sigma^{free}_{NN} ,
\end{equation}
 where $\sigma^{free}_{NN}$ is the experimental nucleon nucleon cross section $[33]$.
The above expression with  $\alpha$ $\approx-0.2$ has been found
to reproduce well the flow data. The free neutron proton cross
section is about a factor of 3 times larger than the free proton
proton or the free neutron neutron one below 400 MeV, which
contributes the main isospin effect from nucleon-nucleon
collisions. We used equation (6) to take into account the medium
effects, in which the neutron proton cross section is always
larger than the  neutron neutron or proton proton cross section in
the medium at the beam energies  in this paper.\\\  
\par
\hskip 0.3in The isospin effect of the in-medium nucleon nucleon
cross section on the observables is defined by the difference
between the observables from an isospin dependent nucleon nucleon
cross section  $\sigma^{iso}$ and that from an isospin independent
nucleon nucleon cross section $\sigma^{noiso}$ in the medium. Here
$\sigma^{iso}$ is defined as $\sigma_{np} \geq
\sigma_{nn}$=$\sigma_{pp}$ and $\sigma^{noiso}$ means
$\sigma_{np}$ = $\sigma_{nn}$ = $\sigma_{pp}$, where
$\sigma_{np}$, $\sigma_{nn}$ and $\sigma_{pp}$ are the neutron
proton, neutron neutron and proton proton cross sections in medium
respectively.
\par
 In order to study the influence played by the medium correction of the
isospin dependent nucleon nucleon cross section on the isospin
effects of the nucleon emission , we investigated the number of
nucleon emission $N_{n}$ as a function of the beam energy for the
mass symmetry neutron-poor system $^{76}K_{r}+^{76}K_{r}$ (top
panels) and mass asymmetry neutron-poor system
$^{112}S_{n}+^{40}C_{a}$ (bottom panels) with the same system mass
at impact parameter b= 4.0 fm for the different symmetry
potentials $U_1^{sym}$ , $U_2^{sym}$ and $U_0^{sym}$ as well as
the isospin dependent in-medium nucleon nucleon cross section
$\sigma^{iso}$ and the isospin independent one $\sigma^{noiso}$
(the explanations about line symbols are in Figures).
 It is clear to see that all of lines with filled symbols are larger than those
with open symbols,i.e.,all of $N_{n}$ with $\sigma^{iso}$ are
larger than those with $\sigma^{noiso}$ because the collision
number with $\sigma^{iso}$ is larger than that with
$\sigma^{noiso}$. We also found that the gaps between lines with
the filled symbols and the open symbols are larger but the
variations among lines in each group are smaller. These mean that
$N_{n}$ depends sensitively on the isospin effect of in-medium
nucleon nucleon cross section and weakly on the symmetry
potential. In particular, the gaps between two group lines with
$\alpha =-0.2$ in left panels are larger than corresponding those
with $\alpha = 0.0$ in right panels, i.e., the medium correction
of two-body collision enhances the dependence of $N_{n}$ on
isospin effect of two-body collision.

To consider the contributions from all of impact parameters we also calculated
an impact parameter average values of nucleon emission
number $<N_{n}>_{b}$ for the same incident channel conditions and
the same symbols as Fig.1.  We can get the same conclusions as Fig.1.\\\
 Why does the medium
correction of nucleon nucleon cross section enhance the dependence
of $N_{n}$ on the isospin effect of two-body collision? As well
know that the isospin dependent in-medium nucleon nucleon cross
section is a sensitive function of the nuclear density
distribution and beam energy as shown in Eq.(6). Fig.2 shows the
time evolution of the ratio of nuclear density to normal one,
$\frac{\rho} {\rho_{0}}$ for the
reaction $^{76}$Kr + $^{76}$Kr and symmetry potential $U_1^{sym}$
at E= 150 MeV/nucleon and b= 4.0 fm. From the values of peak for
$\frac{\rho} {\rho_{0}}$ in the insert in Fig.2 it is clear to see
that  $\rho(\sigma^{iso},\alpha = -0.2)$  is larger
than $\rho(\sigma^{noiso},\alpha = -0.2)$  and
$\rho(\sigma^{iso},\alpha = 0.0)$ is larger than
$\rho(\sigma^{noiso},\alpha = 0.0)$  because the
larger collision number from $\sigma^{iso}$ increases the nuclear
stopping and dissipation , which enhances the nuclear density,
compared to the case with $\sigma^{noiso}$. From Fig.2 we can also
see that $\frac{\rho} {\rho_{0}}$'s decrease quickly with
increasing the time after peak of $\frac{\rho} {\rho_{0}}$.
Because the larger compression produces quick expanding process of
the colliding system ,while the small compression induces slow
expanding process , at the same time, the $\frac{\rho}
{\rho_{0}}$'s decrease quickly with expanding process of system.
Hovewor the decreasing velocity of  $\frac{\rho} {\rho_{0}}$ is larger
for the quick expansion system than that for the slow expansion
system, up to about 70 fm/c, on the contrary,
$\rho(\sigma^{noiso},\alpha = -0.2)$ is larger than
$\rho(\sigma^{iso},\alpha = -0.2)$  and
$\rho(\sigma^{noiso},\alpha = 0.0)$  is larger
than $\rho(\sigma^{iso},\alpha = 0.0)$. In
particular, the gap between two lines for $\alpha = -0.2$ is
larger than that for $\alpha = 0.0$ after about 70 fm/c. This
property is very similar to the $N_{n}$, which means that the
medium correction of an isospin dependent nucleon nucleon cross
section enhances also the dependence of $\rho$ on the isospin
effect of two-body collision, which induces the same effects to
$N_{n}$ through the nucleon nucleon cross section as a function of
the nuclear density as shown in Eg.(6).\\\
We also found an important role of the MDI for enhancing the
isospin effect of two-body collision correlating with the medium
correction of two-body collision on the $N_{n}$ . Fig.3
shows the time evolution of $N_{n}$ for a symmetry potentials
$U_1^{sym}$ at beam energy of 100 MeV/nucleon and impact parameter
of 4 fm. They are four cases (see Fig.3) for MDI in the left
window and NOMDI in the right window. It is clear to see that the
gap between the solid line and dot line with MDI in the left
window is larger than corresponding gap with NOMDI in the right
window in the medium(here only $\alpha = -0.2$), i.e., MDI increases the
isospin effect of two-body collision on the $N_{n}$ in the medium
because above gap is produced from the isospin effect of
nucleon-nucleon cross section in the medium. Here physically there
are two mechanisms at work here. (1) The average momentum of a
particle in medium is higher in a heavy ion collision than in cold
nuclear matter at the same density. (2) MDI induces the
transporting momentum more effectively from one part of the system
to another, in which particles also move with a higher velocity
for a given momentum than in free space.\\
  The degree of isospin fractionation $(N/Z)_{gas}/(N/Z)_{liq}$ is measured by
the ratio of the neutron-proton ratio of the nucleon emission
(gas phase) to that of the intermediate mass fragment
emissions (liquid phase). Where the liquid phase includes all
of fragments with the charge number from 2 to $(Z_{p}+Z_{t})/2$
,where $Z_{p}$ and $Z_{t}$ are the charges for projectile and
target respectively \\\
Fig.4 is plotted the time evolution of the impact parameter
average value of $<(N/Z)_{gas}>_{b}/<(N/Z)_{liq}>_{b}$(top
window), $<(N/Z)_{gas}>_{b}$ (in left side of bottom window) and
$<(N/Z)_{liq}>_{b}$ (in the right side of bottom window) for the
different symmetry potentials $U_3^{sym}$ and $U_1^{sym}$ with the
same $\sigma^{iso}$ or the same $\sigma^{noiso}$ for the
neutron-rich system $^{124}Sn+^{124}Sn$ at the beam energy of 50
MeV/nucleon . It is clear to see that the variations for the
$<(N/Z)_{gas}>_{b}/<(N/Z)_{liq}>_{b}$'s with the $\sigma^{iso}$
and $\sigma^{noiso}$ but the same symmetry potential $U_3^{sym}$
or $U_1^{sym}$ are smaller, but the gap between the
$<(N/Z)_{gas}>_{b}/<(N/Z)_{liq}>_{b}$'s with different symmetry
potentials $U_3^{sym}$ and $U_1^{sym}$ but the same $\sigma^{iso}$
or $\sigma^{noiso}$ are larger. Namely, the amplitude of
$<(N/Z)_{gas}>_{b}/<(N/Z)_{liq}>_{b}$ depends sensitively on the
symmetry potential and weakly on the isospin effect of
in-medium nucleon nucleon cross section.\\\
It is also very clear to see that $<(N/Z)_{gas}>_{b}$ has the same
property as $<(N/Z)_{gas}>_{b}/<(N/Z)_{liq}>_{b}$ but
$<(N/Z)_{liq}>_{b}$ is insensitive to both symmetry
potential and nucleon-nucleon cross section in the medium. Namely above 
property for $<(N/Z)_{n}>_{b}/<(N/Z)_{liq}>_{b}$ is produced mainly from the
gas phase .\\\
What is the main source for producing the isospin fractionation
process in intermediate energy heavy ion collisions?
  As well know that in regions just off normal density the fields
felt  by neutrons and protons are very different for different
symmetry potentials ,especially, below normal density since
symmetry potential is repulsive for neutron and attractive for
proton. We thus expect important transport effects during reaction
at intermediate energies since interacting isospin asymmetry nuclear
matter will experience compressed and expanding phases before
forming fragments around normal density conditions. More complete isospin 
fractionation processes can be obtained from the
analysis of the density dependence of neutron and proton chemical
potentials as above mentioned that the symmetry potential in the chemical 
potential is a repulsive for the
neutron and a attractive for the proton , leading
to the different chemical potential degradations with nuclear
density $\rho $ for the neutron and proton in the low density
region below normal nuclear density and in the expanding process
of the colliding system. However in the nonequilibrium expanding
process the mass flow is determined by the difference in the local
values of the chemical potential and directed from the higher
chemical potential region to lower values until
equilibzation.  Above transport effects induce the unstable isospin
asymmetric nuclear matter to separate into a neutron-rich low
density phase and a neutron-poor high density one. Especially
quite different chemical potential degradations with nuclear
density for neutrons and protons below normal nuclear
density, induce the isospin fractionation ${[30]}$.\\\
In summary, from the calculation results by using IQMD we can get
following conclusions: (1) $N_{n}$ depends sensitively on the
isospin effect of two-body collision and weakly on the symmetry
potential for the neutron-poor system in the higher energy region
. (2) In particular,the medium correction of the nucleon nucleon
cross section enhances the dependence of $N_{n}$ on the isospin
effect of two-body collision .  (3) MDI produces also an important
role for enhancing the dependence of $N_{n}$ on the isospin effect
of two-body collision in the medium. (4) On the contrary ,the
degree of isospin fractionation
$<(N/Z)_{gas}>_{b}/<(N/Z)_{liq}>_{b}$ depends sensitively on the
symmetry potential and weakly on the isospin effect of two-body
collision for neutron-rich system at lower beam energy region.\\
We thank Prof.Bao-An Li for helpful discussions.\\\
This work is supported by the Major State Basic Research
Development Program in China Under Contract No.G2000077400,
"100-person project" of the Chinese Academy of Sciences, The
National Natural Science Foundation of China under Grants
Nos: 10175080, 10004012,10175082 and The CAS Knowledge Innovation Project No. KJCX2-SW-N02\\\

\baselineskip 0.2in
\section*{Figure captions}
\begin{description}
 \item[Fig.1]
  The  nucleon emission number $N_{n}$
as a function of the beam energy for $U_1^{sym}$ , $U_2^{sym}$ and
$U_0^{sym}$ with $\sigma^{iso}$ and $\sigma^{noiso}$ for systems
$^{76}K_{r}+^{76}K_{r}$ and  $^{112}S_{n}+^{40}C_{a}$(see text).
 \item[Fig.2]
The time evolution of the ratio of nuclear density to normal
density $\frac{\rho(\sigma^{iso},\alpha )} {\rho_{0}}$ for four
different cases for system $^{76}Kr+^{76}Kr$ at E=150 MeV/nucleon and b= 4.0 fm (see text).
\item[Fig.3]
 The time evolution of the $ N_{n}$ with MDI and NOMDI for four cases (see text)
and systems $^{76}Kr+^{76}Kr$ at E=100 MeV/nucleon and b= 4.0 fm.
\item [Fig.4]
Time evolutions of an impact parameter average values of the degree
of the isospin fractionation $<(N/Z)_{gas}>_{b}/(<(N/Z)_{liq}>_{b}$ (in top window),
$<(N/Z)_{gas}>_{b}$ and $<(N/Z)_{liq}>_{b}$ (in bottom window) for the  system
$^{124}Sn+^{124}Sn$ at the beam energy of 50 MeV/nucleon and an
impact parameter of 4.0 fm for the different symmetry potentials
$U_1^{sym}$ and $U_3^{sym}$ with $\sigma^{iso}$ and $\sigma^{noiso}$ .

\end{description}

\end{document}